\message{STBASIC.TEX TeX Macro Library}
\message{ }





\def\beginrefs{\begingroup\parindent=0pt\frenchspacing
   \parskip=1pt plus 1pt minus 1pt\interlinepenalty=1000\pretolerance=10000
   \hyphenpenalty=10000\everypar={\hangindent=0.42in}       
  \def\aa##1{{\it Astr.~Ap., \bf ##1}}
  \def\aasup##1{{\it Astr.~Ap.~Suppl., \bf ##1}}
  \def\aasupp##1{{\it Astr.~Ap.~Suppl., \bf ##1}}
  \def\aj##1{{\it A.~J., \bf ##1}}
  \def\annrev##1{{\it Ann.~Rev.\ Astr.~Ap., \bf ##1}}     
  \def\araa##1{{\it Ann.~Rev.\ Astr.~Ap., \bf ##1}}     
  \def\apj##1{{\it Ap.~J., \bf ##1}}     
  \def\apjl##1{{\it Ap.~J. (Letters), \bf ##1}}
  \def\apjlett##1{{\it Ap.~J. (Letters), \bf ##1}}
  \def\apjlet##1{{\it Ap.~J. (Letters), \bf ##1}}
  \def\apjsup##1{{\it Ap.~J.~Suppl., \bf ##1}}
  \def\apjsupp##1{{\it Ap.~J.~Suppl., \bf ##1}}
  \def\baas##1{{\it Bull.~A.A.S., \bf ##1}}
  \def\ban##1{{\it B.A.N., \bf ##1}}
  \def\ibvs##1{{\it Inf. Bull. Var. Stars}, No.~##1}
  \def\mn##1{{\it M.N.R.A.S., \bf ##1}}
  \def\mnras##1{{\it M.N.R.A.S., \bf ##1}}
  \def\pasp##1{{\it Pub.~A.S.P., \bf ##1}}
  \def\ajpasp##1{{\it Pub.~A.S.P., \bf ##1}}
  \def\nat##1{{\it Nature, \bf ##1}}
  \def\nature##1{{\it Nature, \bf ##1}}}

\def\endrefs{\endgroup}



\def\df{\leaders\hbox to 0.6em{\hss.}\hfill}


\def\section#1{\bigbreak\medskip\centerline{#1}\par\nobreak\medskip\markpage}

\def\subsection#1#2{\bigbreak\noindent{\bf#1\hskip 0.9em\relax#2}\par
   \nobreak\medskip\markpage}

\def\subsubsection#1#2{\medbreak\noindent{\sl#1\hskip 0.60em\relax#2}\par
   \nobreak\medskip\markpage}

\def\today{\advance\year by -1900 
   \number\month/\number\day/\number\year}
\def\yearmonthday{\number\year\space
   \ifcase\month\or January\or February\or March\or April\or May\or June\or
   July\or August\or September\or October\or November\or December\fi
   \space\number\day}

\newcount\num

\def\nextnum{\global\advance \num by 1 \number\num}
\def\nextitem{\leavevmode
   \hbox{\ifnum\num>8 \kern-0.43em\fi \nextnum.\kern0.60em}}
\def\bfnextitem{\leavevmode
   \hbox{\ifnum\num>8 \kern-0.43em\fi \bf\nextnum.\kern0.60em}}

\newcount\colnum

\def\nextcolnum{\global\advance \colnum by 1 \number\colnum}
\def\nextcolumn{\leavevmode
   \hbox{{\it \ifnum\colnum<9 \phantom{1}\fi Column \nextcolnum:}\kern0.60em}}

\newcount\fig

\def\nextfig{\global\advance \fig by 1 \number\fig}

\newcount\cap

\def\nextcap{\global\advance \cap by 1 \number\cap}

\newcount\letter

\def\nextlet{\global\advance \letter by 1
   \ifcase\letter\or A\or B\or C\or D\or E\or F\or G\or H\or I\or
   J\or K\or L\or M\or N\or O\or P\or Q\or R\or S\or T\or U\or V\or W\or X\or
   Y\or Z\fi}

\newdimen\bigindent \bigindent=3.5in
\def\letterhead{\hsize=6in\interlinepenalty=2000\parskip=6pt minus 3pt
  \pretolerance=750
  \def\topline##1{\hbox to\hsize{\hfil##1\hskip\rightskip}}
  \footline={\ifnum\pageno=1
    \hss\hbox{\vrule height 0.4in width 0pt}
    \eightrm Operated by the Association of Universities for Research in 
    Astronomy, Inc., for the National Aeronautics and Space Administration\hss
    \else\hfil\fi}
  \null
  \vskip-0.2in
  {\advance\rightskip by -0.75in
    \topline{3700 San Martin Drive}
    \topline{Baltimore, MD 21218}
    \topline{(301) 338-4718}\par}
  \vskip30pt minus 15pt
  {\leftskip=\bigindent\yearmonthday\par}}

\def\arpanetletterhead{\hsize=6in\interlinepenalty=2000\parskip=6pt minus 3pt
  \pretolerance=750
  \def\topline##1{\hbox to\hsize{\hfil##1\hskip\rightskip}}
  \footline={\ifnum\pageno=1
    \hss\hbox{\vrule height 0.4in width 0pt}
    \eightrm Operated by the Association of Universities for Research in 
    Astronomy, Inc., for the National Aeronautics and Space Administration\hss
    \else\hfil\fi}
  \null
  \vskip-0.2in\vskip-3\baselineskip
  {\advance\rightskip by -0.75in
    \topline{3700 San Martin Drive}
    \topline{Baltimore, MD 21218}
    \topline{(301) 338-4718}
    \topline{{\elevenrm BITNET:} \tt golombek@stsci}
    \topline{\elevenrm SPAN: \tt SCIVAX::GOLOMBEK}
    \topline{{\elevenrm ARPANET:} \tt golombek@scivax.arpa}\par}
  \vskip30pt minus 15pt
  {\leftskip=\bigindent\yearmonthday\par}}

\def\gosbletterhead{\hsize=6in\interlinepenalty=2000\parskip=6pt minus 3pt
  \pretolerance=750
  \def\topline##1{\hbox to\hsize{\hfil##1\hskip\rightskip}}
  \footline={\ifnum\pageno=1
    \hss\hbox{\vrule height 0.4in width 0pt}
    \eightrm Operated by the Association of Universities for Research in 
    Astronomy, Inc., for the National Aeronautics and Space Administration\hss
    \else\hfil\fi}
  \null
  \vskip-0.375in
  {\advance\rightskip by -0.75in
    \topline{General Observer Support Branch}
    \topline{3700 San Martin Drive}
    \topline{Baltimore, MD 21218}
    \topline{(301) 338-4996}\par}
  \vskip30pt minus 15pt
  {\leftskip=\bigindent\yearmonthday\par}}



\def\indentleft{\advance\leftskip by 50pt\interlinepenalty=750}
\def\inndentleft{\advance\leftskip by 78pt\interlinepenalty=750}
\def\narrower{\advance\leftskip by 0.42in\advance\rightskip by 0.42in
  \interlinepenalty=750}
\def\nnarrower{\advance\leftskip by 50pt\advance\rightskip by 45pt
  \interlinepenalty=750}

\def\checkbox{\nnarrower\parindent=0pt\itemitem{\vbox{\hrule height.7pt
  \hbox{\vrule width.7pt height6pt \kern6pt \vrule width.7pt}
  \hrule height.7pt}$\,$}}  


%
%
\newcount\index \index=100
\def\markpage{\advance\index by 1 \count\index=\pageno}
\def\begintableofcontents{\begingroup
  \index=100 \frenchspacing\interlinepenalty=750
  \parskip=0.1pt plus 1pt minus 0.1pt \parindent=0.3in
  \def\dfi{\advance\index by 1 \df\number\count\index}
  \def\in{\par\hskip-0.2in\indent \hangindent2\parindent \textindent}    
  \def\inin{\par\hskip0.32in\indent \hangindent3\parindent \textindent}
  \def\ininin{\par\hskip0.95in\indent \hangindent4\parindent \textindent}}



{\obeylines\gdef\startdisplay#1
  {\catcode`\^^M=5$$#1\halign\bgroup\indent##\hfil&&\qquad##\hfil\cr}}
\outer\def\enddisplay{\crcr\egroup$$}

\chardef\other=12

{\obeyspaces\gdef {\ }} 

  \font\twentyfourrm=cmr10 scaled 2488
  \font\twentyfouri=cmmi10 scaled 2074   
  \font\twentyfoursy=cmsy10 scaled 2074
  \font\twentyrm=cmr10 scaled 2074      
  \font\twentyi=cmmi10 scaled 2074   
  \font\twentysy=cmsy10 scaled 2074
  \font\eighteenrm=cmr10 scaled 1728
  \font\eighteeni=cmmi10 scaled 1728 \font\eighteensy=cmsy10 scaled 1728
  \font\fourteenrm=cmr10 scaled 1440
  \font\fourteeni=cmmi10 scaled 1440 \font\fourteensy=cmsy10 scaled 1440
  \font\twelverm=cmr12
                
  \font\twelvei=cmmi12               \font\twelvesy=cmsy10 scaled 1200
  \font\elevenrm=cmr10 scaled 1095
    
  \font\eleveni=cmmi10 scaled 1095   \font\elevensy=cmsy10 scaled 1095
  \font\tenrm=cmr10
                   
  \font\teni=cmmi10  \font\tensy=cmsy10  
  \font\ninerm=cmr9

  \font\ninei=cmmi9                  \font\ninesy=cmsy9
  \font\eightrm=cmr8
  \font\seveni=cmmi7 \font\sevensy=cmsy7

\def\commonstuff{
  \parindent=0.42in       
  \def\skipline{\vskip\baselineskip}
  \hyphenpenalty=200\pretolerance=300\tolerance=600 
  \interlinepenalty=100\clubpenalty=500\widowpenalty=500
  \nonfrenchspacing\singlespace\rm}

\def\twelvepoint{
  \font\bf=cmbx12
  \font\it=cmti12
  \font\sl=cmsl12
  \font\tb=cmtt10 scaled 1200 
  \font\tt=cmtt8 scaled 1440
  \textfont0=\twelverm \scriptfont0=\tenrm     
    \scriptscriptfont0=\sevenrm                 
  \def\rm{\fam0 \twelverm}   
  \textfont1=\twelvei  \scriptfont1=\teni  
    \scriptscriptfont1=\seveni                  
  \def\mit{\fam1 } \def\oldstyle{\fam1 \twelvei}
  \textfont2=\twelvesy \scriptfont2=\tensy 
    \scriptscriptfont2=\sevensy                 
  \def\singlespace{\baselineskip=13.5pt\lineskiplimit=-5pt
    \lineskip=0pt
    \parskip=1.25pt plus 1.5pt minus 0.25pt}  
  \def\oneandahalfspace{\baselineskip=18pt\parskip=0pt plus 1pt}
  \def\doublespace{\baselineskip=24pt\parskip=0pt plus 0.5pt}
  \footline={\ifnum\pageno=1 \hfil
             \else\hss\twelverm-- \folio\ --\hss\fi} 
  \def\pagenumbers{\footline={\hss\twelverm-- \folio\ --\hss}}  
  \def\romanpagenumbers{\footline={\hss\twelverm-- \romannumeral\folio\ --\hss}}
  \commonstuff}

\def\tenpoint{
  \font\it=cmti10
  \font\sl=cmsl10
  \font\bf=cmb10
  \textfont0=\tenrm \scriptfont0=\sevenrm     
    \scriptscriptfont0=\fiverm                 
  \def\rm{\fam0 \tenrm}   
  \textfont1=\teni  \scriptfont1=\seveni  
    \scriptscriptfont1=\fivei                  
  \def\mit{\fam1 } \def\oldstyle{\fam1 \teni}
  \textfont2=\tensy \scriptfont2=\sevensy 
    \scriptscriptfont2=\fivesy                 
  \def\singlespace{\baselineskip=12pt\lineskiplimit=0pt
    \lineskip=-0.5mm       
    \parskip=2pt plus 1pt minus 1pt}  
  \footline={\ifnum\pageno=1 \hfil
             \else\hss\tenrm-- \folio\ --\hss\fi} 
  \def\oneandahalfspace{\baselineskip=18pt\parskip=0pt plus 1pt}
  \def\doublespace{\baselineskip=24pt\parskip=0pt plus 1 pt}
  \def\pagenumbers{\footline={\hss\tenrm-- \folio\ --\hss}}  
  \def\romanpagenumbers{\footline={\hss\tenrm-- \romannumeral\folio\ --\hss}}
  \commonstuff}

\def\elevenpoint{
  \font\it=cmti10 scaled 1095
  \font\sl=cmsl10 scaled 1095
  \font\bf=cmb10 scaled 1095 
  \font\tt=cmtt10 scaled 1095
  \textfont0=\elevenrm \scriptfont0=\tenrm     
    \scriptscriptfont0=\ninerm                 
  \def\rm{\fam0 \elevenrm}   
  \textfont1=\eleveni  \scriptfont1=\teni  
    \scriptscriptfont1=\ninei                  
  \def\mit{\fam1 } \def\oldstyle{\fam1 \eleveni}
  \textfont2=\elevensy \scriptfont2=\tensy 
    \scriptscriptfont2=\ninesy                 
  \def\singlespace{\baselineskip=13pt\lineskiplimit=-5pt
    \lineskip=0mm       
    \parskip=2pt plus 1pt minus 1pt}  
  \footline={\ifnum\pageno=1 \hfil
             \else\hss\elevenrm-- \folio\ --\hss\fi} 
  \def\oneandahalfspace{\baselineskip=19pt\parskip=0pt plus 1pt}
  \def\doublespace{\baselineskip=26pt\parskip=0pt plus 1 pt}
  \def\pagenumbers{\footline={\hss\elevenrm-- \folio\ --\hss}}  
  \def\romanpagenumbers{\footline={\hss\tenrm-- \romannumeral\folio\ --\hss}}
  \commonstuff}

\def\eighteenpoint{           
  \font\bf=cmbx10 scaled 1728
  \font\it=cmti10 scaled 1728
  \font\sl=cmsl10 scaled 1728
  \font\tb=cmtt10 scaled 1728
  \font\tt=cmtt10 scaled 1728
  \textfont0=\eighteenrm \scriptfont0=\fourteenrm
    \scriptscriptfont0=\twelverm                 
  \def\rm{\fam0 \eighteenrm}   
  \textfont1=\eighteeni  \scriptfont1=\fourteeni  
    \scriptscriptfont1=\twelvei                  
  \def\mit{\fam1 } \def\oldstyle{\fam1 \eighteeni}
  \textfont2=\eighteensy \scriptfont2=\fourteensy 
    \scriptscriptfont2=\twelvesy                 
  \def\singlespace{\baselineskip=21pt\lineskiplimit=-5pt
    \lineskip=0pt
    \parskip=4pt plus 1pt minus 1pt}  
  \def\oneandahalfspace{\baselineskip=30pt\parskip=0pt plus 1pt}
  \def\doublespace{\baselineskip=40pt\parskip=0pt plus 1pt}
  \footline={\ifnum\pageno=1 \hfil
             \else\hss\eighteenrm-- \folio\ --\hss\fi} 
  \def\pagenumbers{\footline={\hss\eighteenrm-- \folio\ --\hss}}  
  \commonstuff}

\def\twentypoint{
  \font\bf=cmbx10 scaled 2074
  \font\it=cmti10 scaled 2074
  \font\sl=cmsl10 scaled 2074
  \font\tb=cmtt10 scaled 2074
  \font\tt=cmtt10 scaled 2074
  \textfont0=\twentyrm \scriptfont0=\eighteenrm     
    \scriptscriptfont0=\fourteenrm                 
  \def\rm{\fam0 \twentyrm}   
  \textfont1=\twentyi  \scriptfont1=\eighteeni  
    \scriptscriptfont1=\fourteeni                  
  \def\mit{\fam1 } \def\oldstyle{\fam1 \twentyi}
  \textfont2=\twentysy \scriptfont2=\eighteensy 
    \scriptscriptfont2=\fourteensy                 
  \def\singlespace{\baselineskip=24pt\lineskiplimit=-5pt
    \lineskip=0pt
    \parskip=5pt plus 1.5pt minus 1.5pt}  
  \def\oneandahalfspace{\baselineskip=33pt\parskip=0pt plus 1pt}
  \def\doublespace{\baselineskip=44pt\parskip=0pt plus 0.5pt}
  \footline={\ifnum\pageno=1 \hfil
             \else\hss\twentyrm-- \folio\ --\hss\fi} 
  \def\pagenumbers{\footline={\hss\twentyrm-- \folio\ --\hss}}  
  \def\romanpagenumbers{\footline={\hss\twentyrm-- \romannumeral\folio\ --\hss}}
  \commonstuff}

\def\twentyfourpoint{
  \font\bf=cmbx10 scaled 2488
  \font\it=cmti10 scaled 2488
  \font\sl=cmsl10 scaled 2488
  \font\tb=cmtt10 scaled 2488
  \font\tt=cmtt10 scaled 2488
  \textfont0=\twentyfourrm \scriptfont0=\twentyrm     
    \scriptscriptfont0=\eighteenrm                 
  \def\rm{\fam0 \twentyfourrm}   
  \textfont1=\twentyfouri  \scriptfont1=\twentyi  
    \scriptscriptfont1=\eighteeni                  
  \def\mit{\fam1 } \def\oldstyle{\fam1 \twentyfouri}
  \textfont2=\twentyfoursy \scriptfont2=\twentysy 
    \scriptscriptfont2=\eighteensy                 
  \def\singlespace{\baselineskip=28pt\lineskiplimit=-5pt
    \lineskip=0pt
    \parskip=5pt plus 1.5pt minus 1.5pt}  
  \def\oneandahalfspace{\baselineskip=42pt\parskip=0pt plus 1pt}
  \def\doublespace{\baselineskip=56pt\parskip=0pt plus 0.5pt}
  \footline={\ifnum\pageno=1 \hfil
             \else\hss\twentyfourrm-- \folio\ --\hss\fi} 
  \def\pagenumbers{\footline={\hss\twentyfourrm-- \folio\ --\hss}}  
  \def\romanpagenumbers{\footline={\hss\twentyfourrm-- \romannumeral\folio\ --\hss}}
  \commonstuff}

\def\spose#1{\hbox to 0pt{#1\hss}}
\def\lta{\mathrel{\spose{\lower 3pt\hbox{$\mathchar"218$}}
     \raise 2.0pt\hbox{$\mathchar"13C$}}}
\def\gta{\mathrel{\spose{\lower 3pt\hbox{$\mathchar"218$}}
     \raise 2.0pt\hbox{$\mathchar"13E$}}}

\def\ni{\noindent}
\def\in{\indent}
\def\inin{\in{\in}
\def\ininin{\inin{\in}}}

\twelvepoint
\doublespace
\raggedbottom
\newdimen\sb \def\md{\sb=.01em 
             \ifmmode $\rlap{.}$'$\kern -\sb$
             \else \rlap{.}$'$\kern -\sb\fi}
\newdimen\sa \def\sd{\sa=.1em 
             \ifmmode $\rlap{.}$''$\kern -\sa$
             \else \rlap{.}$''$\kern -\sa\fi}
\null\vskip.5in
\vskip.4in

\centerline{\bf Evolution of the near-infrared luminosity function in
rich galaxy clusters}  
\vskip.4in
\centerline{Neil Trentham}
\smallskip
\centerline{Institute of Astronomy, University of Cambridge}
\vskip 2pt
\centerline{Madingley Road, Cambridge CB3 0HA}
\vskip 8pt
\centerline{and}
\vskip 8pt
\centerline{Bahram Mobasher}
\smallskip
\centerline{Astrophysics Group, Imperial College}
\vskip 2pt
\centerline{Blackett Laboratory, Prince Consort Road, London SW7 2BZ}
\vskip 26pt 
\centerline{Submitted to $MNRAS$}

\vfil
\eject
 
\doublespace
\centerline{\bf ABSTRACT }
\bigskip

\noindent
We present
the $K$-band (2.2 $\mu$) luminosity functions of the X-ray luminous clusters 
MS1054$-$0321 ($z=0.823$), MS0451$-$0305 ($z=0.55$), 
Abell 963 ($z=0.206$), Abell 665 ($z=0.182$) and Abell 1795 ($z=0.063$)
down to absolute magnitudes $M_K = -20$.  
Our measurements probe fainter absolute magnitudes
than do any previous studies of
the near-infrared luminosity function of clusters.
All the clusters are found to have similar luminosity
functions within the errors, when the galaxy
populations are evolved to redshift $z=0$. 
It is known that the most massive bound systems in the Universe
at all redshifts are X-ray luminous clusters.
Therefore, assuming that the clusters in our
sample correspond to a single population
seen at different
redshifts,   
the results here imply that
not only had the stars 
in present-day ellipticals in rich clusters 
formed by $z=0.8$, but
that they existed in as luminous galaxies then as they do today.
 
Addtionally, 
the clusters have $K$-band luminosity functions which appear to be
consistent
with the $K$-band field luminosity function 
in the range
$-24 < M_K < -22$, although the uncertainties in both the field
and cluster samples are large.

\bigskip
\bigskip


\vfil\eject

\noindent{\bf 1 INTRODUCTION}

\noindent
Recent observations of rich clusters of galaxies and their
galaxy populations have revealed a number of
interesting results:
\vskip 4pt
\noindent
(i) Three X-ray luminous clusters at redshifts
$z \sim 0.8$ have been discovered
in the ROSAT North Ecliptic Pole (NEP) survey (Gioia \& Luppino 1994,
Henry et al.~1997), all of which have had their high inferred masses
confirmed by weak gravitational lensing measurements of background
galaxies (Luppino \& Kaiser
1997, Clowe et al.~1998). 
In addition, one cluster at $z=1$ has been
found by ROSAT pointed observations towards the high redshift lensed quasar
MG2016+112 (Hattori et al.~1997).
The very existence of these high-$z$ massive
clusters puts severe constraints on the cosmological geometry
(e.g.~Eke, Cole \& Frenk 1996; Bahcall, Fan \& Cen 1997; Henry 1997). 

\vskip 4pt
\noindent
(ii) Detailed measurments of the optical -- near infrared colours of
early-type galaxies in clusters (Stanford, Eisenhardt \& Dickinson 1998)
suggest that these galaxies evolve passively from $z=0.9$ to $z=0$.
Furthermore, Stanford et al.~provide evidence based on the small
scatter in the optical-IR colours that early-type galaxies in a cluster
show a large degree of homogeneity in their star formation histories,
with very little evidence for uncorrelated recent bursts of star
formation.

\vskip 4pt
\noindent
(iii) Blue star-forming galaxies have been found in a number of rich
clusters at $z > 0.2$.  The existence of such blue galaxies in rich
clusters at intermediate redshifts 
has been recognized for some time (Butcher \& Oemler 1978, 1984), but 
the Hubble Space Telescope ($HST$) now allows these galaxies to be imaged
at very high resolution (Dressler et al.~1994a,b).
Many of these blue galaxies are similar in morphology to nearby
late-type spirals.  Some have disturbed morphologies 
and/or show evidence for mergers.  

\vskip 4pt
\noindent
(iv) Imaging of a sample of $z \sim 0.5$ clusters with WFPC2 on $HST$
suggests that the fraction of elliptical galaxies in the
$z \sim 0.5$ clusters is
similar to or larger than that in local clusters (Dressler et al.~1997). 
However, Dressler et al.~also report that the fraction of
spiral galaxies is much higher in the 
$z \sim 0.5$ clusters. 
A corresponding decrease in the fraction of S0 galaxies
in the $z \sim 0.5$ clusters relative to the fraction in local
clusters is also noted.

\vskip 4pt
\noindent
(v) The $B$-band luminosity function of galaxies in rich, evolved
clusters with high elliptical galaxy fractions is
remarkably invariant (Trentham 1998a).
Compared with the field luminosity function, 
it falls off more steeply at bright magnitudes ($M_B < -20$), 
despite the existence of superluminous cD galaxies in clusters 
not present in the field.
This is probably due to the existence of very massive, star-forming
galaxies in the field (like those seen by Cowie, Hu \& Songalia 1995,
Cowie et al.~1996).
These galaxies do not exist in nearby clusters because 
cluster-related
processes like ram-pressure stripping have turned off star formation
in galaxies there.

\vskip 4pt
\noindent
A consistent picture emerging
from these observations 
is as follows.
The richest clusters were formed quite a long
time ago, and were X-ray luminous by
at least $z=1$.  What we see today as 
the stars in early-type galaxies located in rich clusters
had formed by this redshift, and probably much earlier.  
However, it is not clear 
exactly how many of these stars were 
in a galaxy that was part of an
X-ray luminous cluster at $z=1$.
As the clusters grew, they picked up additional
galaxies, many of which experienced a burst of star formation as they
entered the cluster (see Moore et al.~1996 for a possible mechanism).
These extra galaxies are the spirals 
which were subsequently
converted to S0 galaxies by cluster-related processes, as discussed 
by Dressler et al.~(1997).
Meanwhile, 
the stars in the elliptical galaxies evolved passively.  
By $z=0$, most of the star formation accompanying infall of galaxies into
the cluster 
has finished except in a very few clusters that are merging today.

We now attempt to measure the $K$-band (2.2 $\mu$)
luminosity function for a sample of clusters
with $0 < z < 0.8$. 
The aim here is to explore the evolution of the
shape of the luminosity function 
of the old stellar populations in clusters, 
and its normalization (relative to X-ray luminosity). 
This will allow us to fill in some details in the above scenario. 
For example, 
do the stars in the giant ellipticals
in nearby clusters exist
in such big stellar systems at high redshift clusters, 
or were they in stellar systems
that merged to form the big galaxies we see today?
The observations of the scatter in the
optical -- near-infrared colours (Stanford,
Eisenhardt \& Dickinson 1998) and of the scatter in the UV -- optical
colours of $z \sim 0.5$ 
ellipticals (Ellis et al.~1997) 
indicate that the stars themselves are very old, but do not constrain
the masses or luminosities of the stellar systems in which the stars
were formed.
Also, the present observations will be used to 
determine how the integrated $K$-band luminosity of clusters
(i.e.~the normalization of
the luminosity function) varies with redshift
for clusters of a given
X-ray luminosity (the results of Dressler et al.~1997 suggest that,
at least for the elliptical galaxy population, it does not vary strongly
between $z=0$ and $z=0.5$).  Finally, we will confirm that the 
conclusions of
Stanford, Eisenhardt \& Dickinson (1998) regarding passive
evolution are valid for our sample (we need to do this before making
evolutionary corrections and comparing
luminosity functions between clusters). 
 
At near-infrared wavelengths the light is primarily sensitive to the old
stellar populations (i.e.~the stars that make up the elliptical
galaxies)
with only small
contamination from young star-forming galaxies (like those observed
by Dressler et al.~1984a,b).
Such a study is made feasible by the advent of wide-field
IR arrays on large telescopes so that we can image to faint magnitudes
over wide areas, and reach the faintest galaxies while still having
enough galaxies that the counting statistics are manageable. 
Also, negative K corrections in the $K$-band also help us to reach very
faint absolute magnitudes ($M_K \sim -20$ at $z=0.8$). 

The paper is organized as follows.  In Section 2 we describe our sample
selection and the observations.  We present the near-infrared 
number counts in Section 3
and luminosity functions in Section 4.
In Section 5 we investigate the evolution of luminosity function with
redshift.  Finally we discuss our results in
Section 6.

\vfil \eject

\vskip 18pt
\noindent{\bf 2 SAMPLE SELECTION AND OBSERVATIONS} 

\vskip 8pt
\noindent{\bf 2.1 The sample}

\noindent
The sample used in this study 
consists of six clusters (hereafter MS1054, MS0451,
A41, A665, A963, and A1795) 
in the redshift range $0.06 < z < 0.82$. 
Table 1 lists the cluster richness, redshift $z$, X-ray luminosity 
$L_x$, the Galactic extinction in the $K$-band along our line of sight to the
cluster and the critical surface density $\Sigma_c$ for lensing of
distant background galaxies by 
the cluster dark matter.
The clusters have X-ray luminosities that vary by less than an order
of magnitude.   The
high redshift clusters (MS1054 and MS0451) were selected to have
somewhat higher $L_x$ in order to give 
better counting statistics following background
subtraction.  We were also limited by the lack of known clusters at
$z=0.8$ (only three are known).
In terms of absolute magnitudes, 
these measurements probe somewhat fainter 
(to $M_K \sim -20$) than most previous
studies (e.g.~Barger 
et al.~(1996) reached $M_K = -23$ for their nearest clusters,
and Stanford, Eisenhardt \& Dickinson (1995) reached $M_K = -22$ in their
study of A370 and A851). 
 
We assume $H_0$ = 75 km s$^{-1}$ Mpc$^{-1}$ and
$\Omega_{0} = 1$ throughout this work. 

\vskip 8pt
\noindent{\bf 2.2 The $K$-band observations and data reduction}

\noindent
Our observing log is presented in Table 2 which lists
the field sizes, coordinates, total exposure times
and seeing.

The clusters MS1054, MS0451, and A41 were imaged through a $K$ filter
using the IRCAM3
256 $\times$ 256 InSb array (scale 0.286$^{\prime \prime}$ pixel$^{-1}$,
field of view $1.2^{\prime} \times 1.2^{\prime}$) 
at the 3.8 m United Kingdom Infrared
Telescope (UKIRT) on Mauna Kea.     
Conditions were photometric with a median seeing of 
0.8$^{\prime \prime}$.
The frames were taken in sequences of nine, one-minute exposures. 
The exposures were dithered by up to 10$^{\prime \prime}$ so
that we could use them to 
reject bad pixels and generate sky frames (the largest objects in
the field for these distant clusters were smaller than this).

The clusters A963, A665, and A1795 were imaged through a $K^{\prime}$
filter using the QUIRC
1024 $\times$ 1024 InSb array (Hodapp et al.~1996,
scale 0.19$^{\prime \prime}$ pixel$^{-1}$,
field of view $3.2^{\prime} \times 3.2^{\prime}$) 
at the 2.2 m University of Hawaii Telescope, also 
on Mauna Kea.
Conditions were photometric for these observations with a median seeing
of about 0.9$^{\prime \prime}$, 
except for A1795 where the seeing was approximately
1.3$^{\prime \prime}$ due to severe windshake of the
telescope. 
Exposures were taken in sequences of six, three-minute frames.
These images were dithered by up to 30$^{\prime \prime}$ in order to
reject bad pixels.  Offset fields were observed with
equal exposure time to the
cluster fields in order to perform a sky subtraction.
The cD galaxies in these fields were too large to allow the use of 
the object frames to create a sky image for subtraction.

Flat-field images were constructed using either median-filtered object 
frames from the whole night (IRCAM3) 
or dome flats (QUIRC), and sky images were constructed
for each exposure sequence using either the object frames (IRCAM3) or
offset blank sky frames (QUIRC).
Individual frames were flat-fielded, sky-subtracted, and then registered
and combined.  The reduced images were flat to better than 1\%.
When combining images a clipping algorithm was used,
which rejected pixels more than 3 standard deviations from the median
value (this removed bad pixels).  
Instrumental magnitudes were computed from observations of
5 $-$ 10 UKIRT faint standards (Casali \& Hawarden 1992)
per night, giving a photometric accuracy of $<2$ \%.
The final $K$-band images are presented in Figure 1.
 
For the QUIRC images, a correction from $K^{\prime}$ to $K$ magnitudes 
was made using the colour calibrations of Wainsocat \& Cowie (1992),
assuming $H-K \approx 0.75$ (this is appropriate for early-type
galaxies with $0.06 < z < 0.2$ -- see Stanford, Eisenhardt \&
Dickinson 1995).  This method gives the correct $K$ magnitudes for
the cluster ellipticals (which are the dominant galaxies in these
clusters), 
although it gives magnitudes that are
systematically too bright 
for bluer background/foreground galaxies and stars, by up to 0.1 mag.

\vfil \eject
\vskip 8pt
\noindent{\bf 2.3 Optical data and measurment of colours}

\noindent
Part of the aim of this project is to compare galaxies in different
clusters at different redshift.  We expect to be able to do this because
of the results of Stanford, Eisenhardt \& Dickinson (1998) who find
that the colours and magnitudes of
elliptical galaxies in clusters vary in a manner
consistent with passive evolution.
However, before we can compare clusters we need to confirm that the galaxies
in our clusters do indeed follow the evolutionary tracks described by
Stanford et al.~(1998).  In order to do this we need to measure
optical-$K$ colours for the galaxies, and so need optical images of the
clusters.

For A665 and A963, we used the $R$-band
data of Trentham (1998b) -- note that the
A665 image in that paper does not completely
overlap with the field studied in the
present work, but a shorter (5 min) exposure taken under the same
conditions as the deeper exposures, but encompassing all the
field studied here, was used.
For A1795, we used the $R$-band data of Trentham (1997b).
The observations and reduction of the optical data is described in those
papers.
For MS1054 and MS0451, we used HST archive data taken using the
F814W and F702W filters respectively and WFPC2.
The exposure times were 260 miutes for MS1054 and
173.3 minutes for MS0451.
We reduced and calibrated these data
using standard procedures -- cosmic rays were identified by their
presence in only one of the dithered set of images and 
subsequently removed.

Colours were measured using 3$^{\prime \prime}$ 
diameter apertures
placed at the same position in registered optical and $K$-band images.
Using aperture magnitudes to measure colours ensures that we
are studying the same
stellar populations in both
filters.  The large aperture ensures that differential seeing
effects between the two images are negligible.

\vskip 18pt
\noindent{\bf 3 THE $K$-BAND GALAXY COUNTS} 

\noindent

Objects were detected at the 3$\sigma$ level above the sky background
in each of the six cluster images, using the FOCAS detection
algorithm (Jarvis \& Tyson 1981; Valdes 1982, 1989).   
Both $3 \sigma$ isophotal magnitudes and magnitudes within a
3$^{\prime \prime}$ diameter aperture were calculated.

The detection algorithm produced catalogs of
100 -- 200 objects per cluster brighter than
the limiting magnitude 
($K \approx 19$ for A1795 and A41, $K \approx 20$ for A665 and A963,
$K \approx 21$ for MS1054 and MS0451).   Each object was then  
examined and compared with objects at the same
($\alpha$, $\delta$) in deeper optical images (see Section 2.3).
At this stage a number of alterations were made to the catalog.
\vskip 4pt
\noindent
(i) The
brightest objects ($K<16$ in the QUIRC images, $K < 17$ in the A41 image,
and $K < 19$ in the deep IRCAM3 MS1054 and MS0451 images) were identified
as stars or galaxies, based on their morphologies, using the PSF-fitting
algorithm DAOPHOT (Stetson 1987).
\vskip 4pt
\noindent
(ii) Multiple objects were identified either 
by their morphologies in the optical
images or by using the FOCAS detection algorithms to search for multiple
peaks within a detection isophote in the $K$-band images.  
A multiple object identified by either
of these methods was then split into its component objects and photometry
performed on each of these separately. 
\vskip 4pt
\noindent
(iii) Objects fainter than the point-source limiting magnitudes given
above were
removed.  For these distant clusters, the early-type galaxies are compact
and the fainter ones
have scale-lengths smaller than the seeing, so that the completeness limit
for these galaxies will be similar to that for point sources.
The few objects fainter than this limit in our catalog arise from 
faint galaxies being superimposed on noise peaks (these galaxies would
not have been detected if they had fallen in a region of the image other
than on a noise peak).
\vskip 4pt
\noindent 
It was also apparent from comparing the $K$-band and optical images
that the effects of crowding (the process by which a faint object goes
undetected because it happens to fall within the detection isophote of
a much brighter object) were small in the
$K$-band images, and can safely be neglected when computing the number
counts.  This is not true for optical CCD images where a bright 
star or galaxy
is present because the diffraction spikes or diffuse halo
can cover considerable
area. 

In order to measure the galaxy counts we need to determine how many of the
detected objects are stars.
Stars
brighter than the magnitudes given in (i) above were identified
by their morphology in either the $K$-band or optical images.
Because the faintest detected
galaxies have scale-lengths 
smaller than the seeing, they look like stars.  Hence, we could not 
identify stars unambiguously based on their morphology at faint magnitudes.
Therefore, we corrected for stellar contamination at faint magnitudes
by computing the expected number of faint stars, given the number of
stars brighter than the limits in (i) above, assuming the Galactic
stellar number-count vs.~magnitude relation slope 
of Jones et al.~(1991).  
These corrections were small ($< 5$\% for all clusters except for
A665, where the stars comprise approximately 20\% of the total counts
at the faintest magnitudes), implying that the uncertainties generated by
them in the context of computing luminosity functions are 
negligible compared to those from counting statistics
and the field-to-field variance in the background. 
Corrections for Galactic extinction in the $K$-band
along the lines of sight to these clusters
are also small ($\leq 0.01$ mag). 
The corrections listed in Table 1 were applied
to the magnitudes of individual objects in our catalog.

Simulations 
of distant compact galaxies indicate that
the $3 \sigma$ isophotal magnitudes
($\sim 21.5 K$ mag arcsec$^{-2}$ for the QUIRC images and for A41,
and $\sim 22.5 K$ mag arcsec$^{-2}$ for the IRCAM3 images of MS1054 and
MS0451) are close to the total magnitudes (Trentham 1997a), and we 
assume this to be the case for all the galaxies here.
Making this assumption, we bin the data in one-magnitude intervals,
correct for stellar contamination as above, and generate the 
galaxy number count $-$ magnitude relation for each of the cluster
fields.  These are presented in Fig.~2, where  
we also show the average background counts, computed from the compilation
in Mobasher \& Trentham (1998).  

It is clearly visible from Figure 2 that for all clusters except A41, the
galaxy number counts significantly exceed the 
background.  We can now subtract the background from these counts
and compute the luminosity functions (Section 4).  
We do not consider A41 further because the cluster is not visible above
the background at any magnitudes fainter than that of its cD galaxy.
It is the least X-ray luminous cluster in our sample, and is the third most
distant so it is not surprising that the background contamination is
the worst here. 

\vskip 18pt 
\noindent{\bf 4 THE LUMINOSITY FUNCTIONS} 

\noindent
The luminosity functions are computed by subtracting the background
(the dashed lines in Figure 2) from the number counts in each cluster
field (the points in
Figure 2).  Additional uncertainty is generated by the fact that 
background fields of a given angular size have a substantial field-to-field
variance (see Figure 3 in Mobasher \& Trentham 1998).   Poisson statistics
are used to correct for the different area in the fields.

The background contribution and its variance 
is further multiplied by a 
correction factor that takes into account the gravitational lensing
of the background galaxies by the cluster dark matter (see Fig.~3).
We need to make this correction because we are looking at the background
galaxies through a large concentration of dark matter 
which will distort their fluxes and angular separations due to
gravitational lensing.
The correction is the
largest for MS0451.  For MS1054, the cluster is sufficiently
distant that many faint non-cluster members are foreground galaxies so that
the average amplification per galaxy is low. 
For the lower
redshift clusters, the increase in angular spacing between galaxies 
due to gravitational lensing progressively becomes more important
compared to the increase
in flux of the galaxies, so 
that the average amplification per galaxy is also low.
For a more detailed description of the shapes of the curves in Fig.~3,
the reader is referred to Trentham (1998b). 

The luminosity functions
are presented as a function of the absolute magnitude plus K correction
(Fig.~4);
this is the most general way of presenting the luminosity functions 
because
the K corrections are slightly different for different galaxy types. 
The error bars here
represent the quadrature sum of the uncertainties from counting statistics
and uncertainties from the field-to-field variance of the background
galaxies.
The luminosity functions have approximately the same shape for all the 
clusters (the normalizations depend primarily on the area of the cluster
surveyed -- see Table 3); however, the error bars are large.    
Before performing a detailed comparison between
the luminosity functions, we need to
correct the galaxy magnitudes for evolutionary effects, as well as make
appropriate K corrections.  This is the subject of the next section. 

\vskip 18pt 
\noindent{\bf 5 EVOLUTION OF
THE NEAR-INFRARED LUMINOSITY FUNCTION}

\noindent
Figure 5 presents the optical-$K$
colour-magnitude diagrams for individual clusters.
The elliptical galaxy sequence is clearly visible in each of
the panels; the slight tendency towards bluer colours at fainter
magnitudes is due to a decrease in metallicity (see e.g.~Bower,
Lucey \& Ellis 1992).
The median colour of galaxies within 3 magnitudes of the brightest
galaxy detected in each cluster (this is the cD for all but A1795)
is presented in Table 4 where it is compared with 
the predicted colour for
passively evolving luminous
ellipticals at the redshift of the cluster.  The
table shows that the two numbers agree very well for all our clusters 
which is not surprising in light of the results of Stanford, Eisenhardt
\& Dickinson (1998), as
described in Section 1.  This implies that we 
can now make joint evolutionary (which arise because the spectral
energy distributions of the galaxies vary with cosmological
time) + K (which arise because the $K$-band is probing
different regions of rest-frame wavelength space for galaxies in
clusters seen at different redshifts) corrections to
the cluster luminosity functions directly from stellar population
synthesis models. 
We can then compare galaxies in clusters at different redshifts.

We estimate these corrections using the 
prescription of Pozzetti, Bruzual \& 
Zamorani
(1996) which is based on the population synthesis models of
Bruzual \& Charlot (1993) -- see Table 5.
In effect what is done is to evolve a local old stellar
population backward in time and derive 
the corrections for the $K$-band at each cluster redshift.  We
then apply the corrections to the LFs and correct them to $z=0$.
We apply the correction to the LF as a whole, assuming corrections
appropriate to E/SO galaxies, and not to each
galaxy individually since we are unable to measure the morphologies
and Hubble types of the galaxies in the distant clusters because
the scale-lengths are smaller than the seeing (for the nearby clusters
the $K$-band evolutionary + K corrections only vary weakly with Hubble
type; see Fig.~2(b) of Pozzetti et al.~1996).  
That the E/S0 approximation is a good one is suggested by the
results of Stanford, Eisenhardt
\& Dickinson (1998) and confirmed for the clusters in our sample
by the colours that
we measure, as presented in Figure 5 and described in the previous
paragraph.  

A weighted average of luminosity functions of these five
individual clusters is
computed and presented in Fig.~6. 
The $K$-band field luminosity function from a recent medium-deep survey 
(Szokoly et al.~1998) is
presented there as well, normalized to minimize scatter with the
cluster points.

The luminosity functions for individual clusters are
all consistent with this composite
function within a normalization constant
(see Table 6 for the normalization constants and parameters that describe
the quality of the fits). 
The agreement between the individual clusters and the composite function
appears to be very good, but the statistics
for each individual cluster are very poor (particularly for MS1054).
However,
within the uncertainties, it appears that the shape of the
$K$-band luminosity function in rich clusters does not vary
strongly with redshift.

A similarly close agreement was found
between the luminosity functions of different
rich clusters in the $B$-band (Trentham 1998a), where the
statistics were much better.
The average slopes of the luminosity function between $M_{cD} + 1.5$
and $M_{cD} + 4.5$ is $\alpha = -1.33 \pm 0.07$ in the $B$-band and
$\alpha = -1.38 \pm 0.24$ in the $K$-band
(here $M_{cD}$ is the magnitude of the brightest cD galaxy
in the sample and $\alpha$ is the logarithmic slope of the luminosity
function: $\log N = -0.4 (\alpha + 1 ) M +$ constant, by convention). 
The concordance between these two numbers is not surprising because
both luminosity functions are probing the same galaxy population
(i.e.~the giant 
ellipticals) at bright magnitudes -- in Abell 963, which is
a Butcher-Oemler cluster, the blue population is $<20$ \% of the total
galaxy counts at bright magnitudes.
 
We also note a concordance between the composite cluster
and field luminosity functions in the $K$-band, although this may in
part be due to poor statistics.
The significant difference observed between the Trentham (1998a)
cluster and Loveday et al.~(1992) field samples
at the very bright end was only seen at $M_B < -20$.  In the $K$-band, 
this corresponds to $M_K < -24$.  The luminosity
functions presented in this work have error bars that
are too large to assess 
whether such a difference is present in the $K$-band (if the
interpretation given in Section 1 of this paper is correct, we would 
expect the difference to be significantly smaller in $K$ than in $B$).
 
Therefore, the numbers in Table 6 suggest that
the shape of the luminosity function is consistent within
our sample.  In Fig.~7 we investigate how the normalization varies.
Here we plot the projected $K$-band luminosity density of the cluster
relative to the total X-ray luminosity.
The figure has large error bars, but this ratio does not appear to
systematically vary with redshift.
The likelihood here
is that the total $K$-band luminosity of a rich cluster
of given X-ray luminosity at any redshift, 
when passively evolved to $z=0$, does 
not depend strongly on the redshift of that cluster.
For this to be true, we need to assume that the clusters in this
sample are representative for rich X-ray clusters at their 
respective redshift.  While
this assumption is fairly secure for $z \sim 0.2$, there are hints that it
may not be correct at higher redshifts.  This is addressed in 
the next section.  
As more deep $K$-band images of galaxy clusters 
become available, better statistics
will be acquired for the kind of comparison shown in Fig.~7
(the other studies
that we list in Section 1 are not deep enough to be placed on this figure);
these observations
will provide a strong constraint on both the formation and evolution of old
stellar populations in clusters. 

\vskip 18pt
\vfil\eject
\noindent{\bf 6 DISCUSSION AND CONCLUSIONS}

\noindent
The main result of this study
is that the shape and normalization (relative to the total X-ray
luminosity of the cluster) of
the $K$-band luminosity functions of X-ray luminous clusters do not
change between $z = 0.2$ and $z=0.8$ (the redshift of the most
distant known X-ray luminous clusters), within the errors of our
measurements. 
One cannot be certain that
we are probing the same population of objects seen at
different redshifts because the X-ray luminosities 
of specific objects in a given X-ray energy band
may increase with time because of accretion of gas  
or decrease because of evaporation.
However, at all redshifts, these X-ray luminous clusters are
the largest bound objects known in
the Universe.
Therefore, the indications are that by looking at 
galaxies in such objects at different redshifts, we are  
probing the evolution of galaxies in a single population of clusters.
The fact that the luminosity function does not change significantly
suggests that neither tidal destruction (which would skew the
luminosity function towards lower luminosities at lower redshifts) or
merging of galaxies (which would skew the luminosity function towards
higher luminosities at lower redshifts) are important processes for
early-type galaxies in clusters between $z=0.8$ and $z=0.2$.
Presumably tidal destruction is not an efficient process because the
early-type galaxies are dense (this is inferred from their position in
the infrared fundamental planes -- Pahre, Djorgovski \& De
Carvalho 1997, Mobasher et al.~1998) and not easily
destroyed. 
Also, merging is not expected to be
an efficient process because the clusters
have high velocity dispersions and hence galaxies are moving fast relative
to each other. 
When combined with the results of Stanford, Eisenhardt \& Dickinson
(1998), our results suggest that not only had the stars 
seen in present-day cluster ellipticals formed by $z=0.8$, but
they existed in as luminous galaxies then as they do today.

The $K$-band luminosity functions are 
very insensitive to contributions from 
late-type galaxies that are currently
forming stars and are very luminous
in the $B$-band. 
As explained in Section 1, such galaxies do exist in clusters at
high redshifts, but 
they are probably a different population 
of galaxies altogether from the present-day cluster ellipticals (they
are more likely to be associated with present-day cluster lenticular
galaxies -- Dressler et al.~1997).

The main limitations of this work are as follows.
Firstly, the sample is small and the error bars in Figs.~6 and 7 are 
quite large.  Obtaining better statistics at the faint end is very
difficult because the background contamination is severe.  Better
statistics at the bright-end are difficult to obtain because we need
to image over wide areas to improve counting statistics.  Such measurements
will be much easier with the next generation of 2K and bigger mosaic
infrared arrays that are currently under construction.
With larger arrays we would also be able to extend our analysis down
to $z=0$. 
Secondly, it is not clear that we really are probing typical examples
of the most X-ray luminous objects in the Universe at high redshift.
If optically dark high redshift clusters
(the MG2016+112 lensing cluster may be an example) are common, those
at $z>1$ 
would be undetected in the ROSAT NEP survey, and those at $z \sim 1$ which
are 
detected with ROSAT would be difficult to identify as a galaxy cluster
based on optical follow-up work. 
These would have very different galaxy populations from clusters like
MS1054.  If MS1054 is atypical for an X-ray luminous cluster at high
redshift, this would weaken or invalidate the above conclusions. 
 
Finally, the clusters have a $K$-band luminosity function which is consistent
with the $K$-band field luminosity function of Szokoly et al.~(1998) for
$-24 < M_K < -22$.
We cannot make a comparison brighter than this because of poor
counting statistics in the cluster sample, or fainter than
this because the field sample does not reach such faint limits. 
 
\vskip 18pt
\noindent{\bf ACKNOWLEDGMENTS}

\noindent We are grateful to Tim Hawarden
for taking the UKIRT observations of MS1054, MS0451 and A41 in December,
1997.  NT acknowledges the PPARC for financial support. 

\vskip 18pt
\ni{\bf REFERENCES }

\beginrefs

Abell G.~O., 1958, ApJS, 3, 211

Bahcall N.~A., Fan X., Cen R., 1997 ApJ, 485, L53

Barger A.~J., Arag{\'o}n-Salamanca A., Ellis R.~S., Couch W.~J.,
Smail I., Sharples R.~M., 1996, MNRAS, 279, 1

Bower R.~G., Lucey J.~R., Ellis R.~S., 1992, MNRAS, 254, 601

Bruzual A.~G., Charlot S., 1993, ApJ, 405, 538 

Burstein D., Heiles C., 1982, AJ, 87, 1165

Butcher H.~R., Oemler A., 1978, ApJ, 219, 18

Butcher H., Oemler A., 1984, ApJ, 285, 426  

Cardelli J.~A., Clayton G.~C., Mathis J.~S., ApJ, 345, 245

Casali M.~M, Hawarden T.~G., 1992, The JCMT-UKIRT Newsletter Vol.~4, p.~33

Clowe D., Luppino G.~A., Kaiser N., Henry J.~P., Gioia I.~M., 1998,
ApJL, in press  

Cowie L.~L., Hu E.~M., Songaila A., 1995, Nat, 377, 603

Cowie L.~L. Songaila A., Hu E.~M., Cohen J.~G., 1996, AJ, 112, 839

Dressler A., Oemler A., Butcher H.~R., Gunn J.~E., 1994a, ApJ, 430, 107 

Dressler A., Oemler A., Sparks W.~B., Lucas R.~A., 1994b, ApJ, 435, L23 

Dressler A., Oemler A., Couch W.~J, Smail I., Ellis R.~S., Barger A.,
Butcher H., Poggianti B.~M., Sharples R., 1997, ApJ, 490, 577 

Eke V.~R., Cole S., Frenk C.~S., 1996, MNRAS, 282, 263

Ellis R.~S., Smail I., Dressler A., Couch W.~J., Oemler A., Butcher H.,
Sharples R,~M., 1997, ApJ, 483, 582

Gioia I.~M., Luppino G.~A., 1994, ApJS, 94, 583

Girardi M., Fadda D., Giuricin G., Mardirossian F., Mezetti M.,
Biviano A., 1996, ApJ, 457, 61

Hattori M.~et al., 1997, Nat, 388, 146 

Henry J.~P., 1997, ApJ, 489, L1

Henry J.~P.~et al., 1997, AJ, 114, 1293

Hodapp K.~W.~et al.~1996, New Astron., 1, 77  

Jarvis J.~F., Tyson J.~A., 1981, AJ, 86, 476 

Jones C., Forman W., 1984, ApJ, 276, 38

Jones L.~R., Fong R., Shanks T., Ellis R.~S., Peterson B.~A., 1991,
MNRAS, 249, 481.

Loveday J., Peterson B.~A., Efstathiou G., Maddox S., 1992, ApJ, 390, 338

Luppino G.~A., Kaiser N., 1997, ApJ, 475, 20

Mobasher B., Guzm{\'a}n R., Arag{\'o}n-Salamanca A., Zepf S., 1998, MNRAS, 
in press 

Mobasher B., Trentham N., 1998, MNRAS, 293, 315 

Moore B., Katz N., Lake G., Dressler A., Oemler A., 1996, Nat, 379, 613

Mushotzky R.~F., 1984, Physica Scripta, T7, 157 

Pahre M.~A., Djorgovski S.~G., De Carvalho R.~R., 1995, ApJ, 453, L17

Pozzetti L., Bruzual A.~G., Zamorani G., 1996, MNRAS, 281, 953

Quintana H., Melnick J., 1982, AJ, 87, 972

Sarazin C.~L., 1986, Rev.~Mod.~Phys., 58, 1

Soltan A., Henry J.~P., 1983, ApJ, 271, 442

Stanford S.~A., Eisenhardt P.~R.~M., Dickinson M., 1995, ApJ, 450, 512

Stanford S.~A., Eisenhardt P.~R., Dickinson M., 1998, ApJ, 492, 461

Steston P.~B., 1987, PASP, 99, 191

Szokoly G.~P., Subbarao M.~U., Connolly A.~J., Mobasher B., 
1998, ApJ, 492, 452

Trentham N., 1997a, MNRAS, 286, 133

Trentham N., 1997b, MNRAS, 290, 334  

Trentham N., 1998a, MNRAS, 294, 193 

Trentham N., 1998b, MNRAS, 295, 360 

Valdes F., 1982, Proc.~SPIE, 331, 465 

Valdes F., 1989, in Grosbol P.~J., Murtagh F., Warmels R.~H., ed.,
ESO Conference and Workshop Proceedings No.~31:
Proceedings of the 1st ESO/St-ECF Data Analysis Workshop.
European Space Observatory, Munich,  p.~35

Wainscoat R.~J., Cowie L.~L., 1992, AJ, 103, 332

\endrefs

\vfil
\eject

\ni {\bf FIGURE CAPTIONS}
\medskip
\bigskip

\noindent {\bf Figure 1.~}The $K$-band images of the clusters fields
used in this study.
In each image North is up and East is to the left.
The images were reduced as described in the text; instrumental
configurations, exposure times, and field sizes are given in Table 2.

\vskip 8pt
\noindent {\bf Figure 2.~}
Galaxy number counts versus $K$ magnitude for the sample clusters.
The clusters which were observed through the $K^{\prime}$ filter have
their magnitudes converted to $K$ magnitudes using the conversion
equation $K^{\prime} - K = 0.22 (H-K)$ (Wainscoat \& Cowie 1992), assuming 
$H-K \approx 0.75$ (Stanford, Eisenhardt \& Dickinson 1995) for typical
cluster galaxies at $0.06 < z  < 0.2$, most of which are early-type.
The dashed lines show the predicted background counts (from the compilation
of Mobasher \& Trentham 1998).  The error bars come from counting statistics. 

\vskip 8pt
\noindent {\bf Figure 3.~}
Corrections for
gravitational lensing applied to the background counts before
subtraction, for each of the cluster fields.  The ratio $f_{\rm lens}$
represents the ratio of background/foreground number counts for a line of sight
seen through the cluster dark matter relative to the background/foreground
number counts in a typical random blank field.  The results here were
calculated according to the recipe given
in Trentham 1998b, assuming isothermal
sphere lenses with velocity dispersions $\sigma$ 
of 1210 km s$^{-1}$ for MS1054,
1510 km s$^{-1}$ for MS0451, 1290 km s$^{-1}$ for A963,
1200 km s$^{-1}$ for A665 and 887 km s$^{-1}$ for A1795. 
The velocity dispersion for A1795 is that measured by Girardi
et al.~1996); the others were
derived using the $L_{x} - \sigma$ correlation of Quintana \& Melnick (1982). 

\vskip 8pt
\noindent {\bf Figure 4.~}
Luminosity functions for the sample clusters, computed from the number
counts (Fig.~2) as described in the text.  The luminosity function of
A41 is not presented since the background
counts dominate at all magnitudes fainter than that of the cD galaxy.
The error bars here represent the quadrature sum of errors from counting
statistics and the field-to-field variance in the background.

\vskip 8pt
\noindent {\bf Figure 5.~}
Colour-magnitude diagrams for the cluster fields.
The colours are those for a 3 arcsecond diameter aperture as described in
the text.  This aperture is large enough ($>2$ FWHM for A1795, and
$>3$ FWHM for the other clusters) that systematic effects due to 
differences in the seeing between the optical and $K$-band images
are small.  The colours for MS1054 and MS0451 assume ST magnitudes
(see e.g.~Synphot Users Guide published by STSci) 

\vskip 8pt
\noindent {\bf Figure 6.~}
The composite luminosity function.  This is the weighted average of
the combined evolution- and K- corrected luminosity functions of the individual
clusters, where the corrections are made as described in the text.
Here $M_{K, corr}$ represents the absolute $K$-band magnitude of a galaxy
in a distant cluster passively evolved to $z=0$.   
The normalization is arbitrarily selected to be for a cluster having
the same number of galaxies as A963 in the $-23 < M_{K, corr} < -22$ bin.
The field points are from the data of Szokoly et al.~(1998; median
sample redshift $z \sim 0.2$), evolved
to $z=0$ in a manner consistent with the cluster data.
This was normalized so as to minimize the scatter with the cluster
luminosity function. 

\vskip 8pt
\noindent {\bf Figure 7.~}
The integrated $K$-band luminosity per square kpc 
in galaxies brighter than $M_{K, corr} = -20$, relative
to the X-ray (2 $-$ 10 keV) luminosity for our sample,
plotted as a function of redshift.   
Presenting the $K$-band total luminosities as a density allows us
to correct for the (small) areas in our cluster field
rest-frame sizes, while maintaing as small error bars as
possible.  In all cases the field size corresponds approximately
to the cluster core (see Table 3).
We do not include A1795 on this plot since only a small
region of the core is imaged there which was selected {\it a priori} to have
a higher than average density of elliptical galaxies.
The X-ray luminosities were computed from the luminosities in Table 1,
assuming a thermal brehmsstrahlung spectrum (neglecting the frequency
dependance of the Gaunt factors -- see Sarazin 1986) 
and the temperature-luminosity relation of Mushotzky (1984). 

\par\vfill\eject
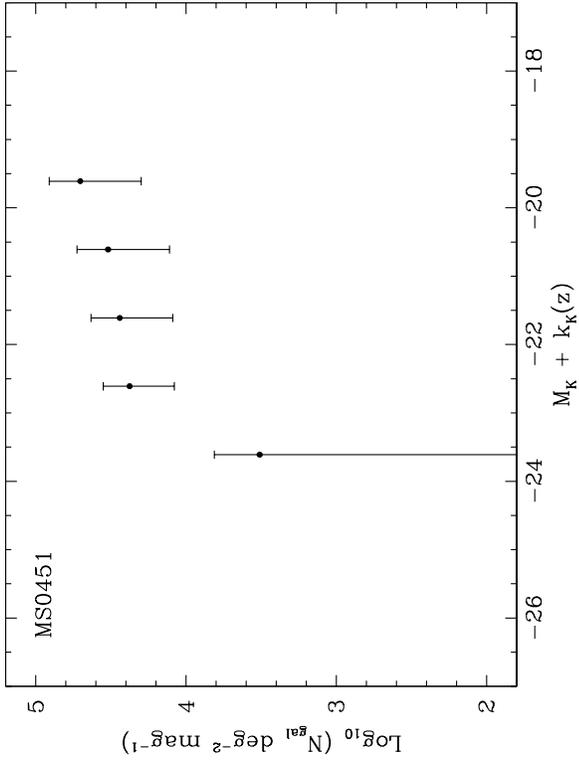
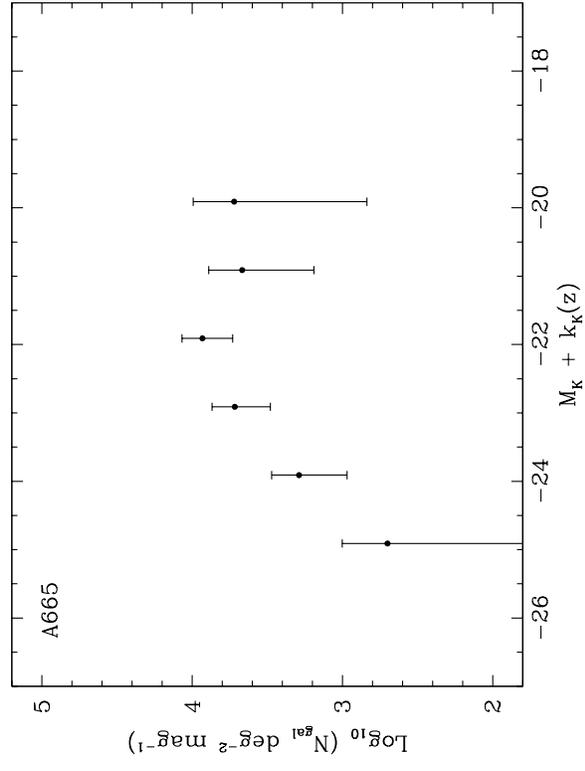
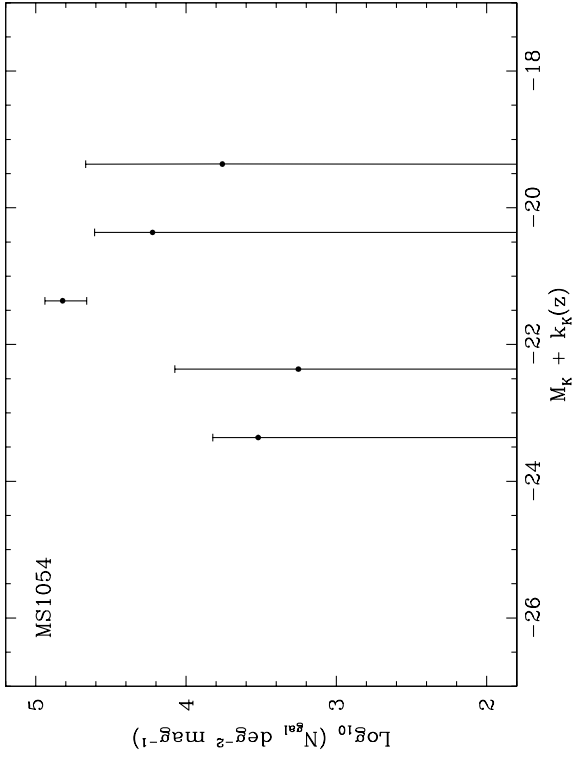
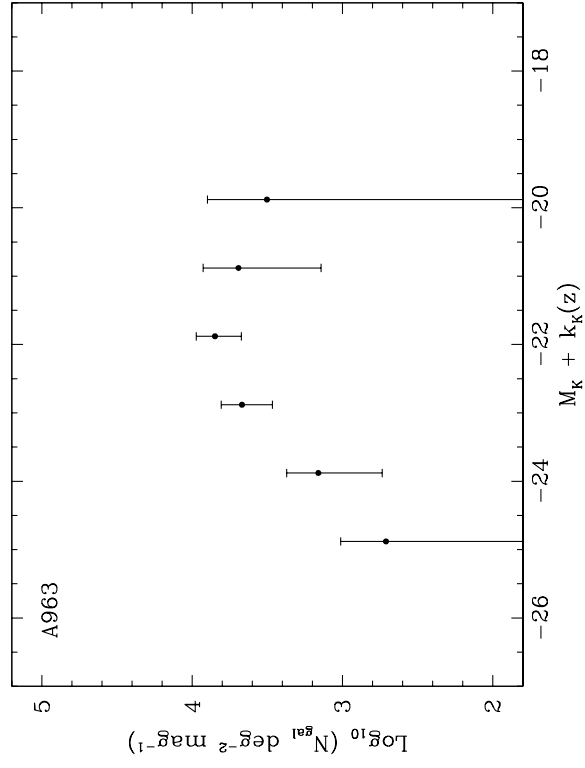
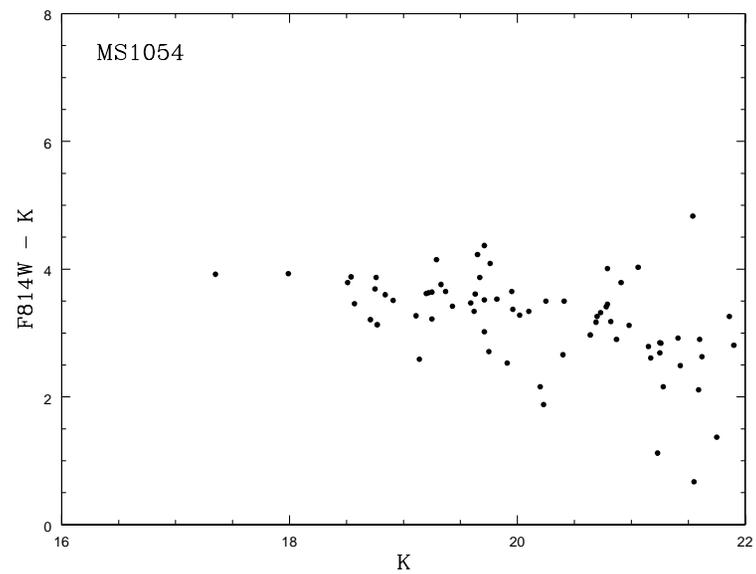
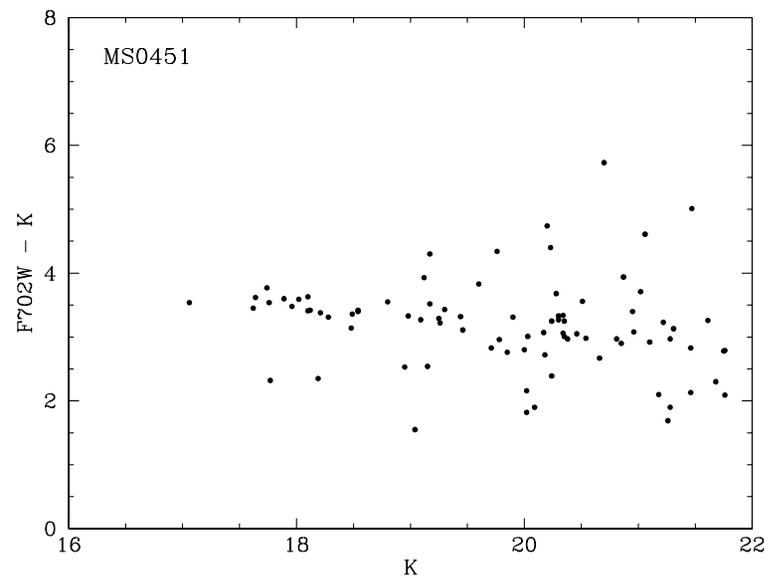
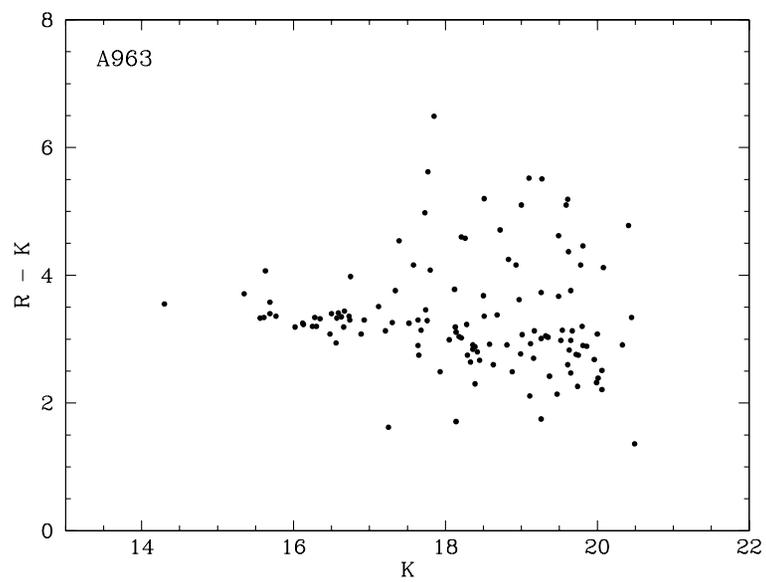
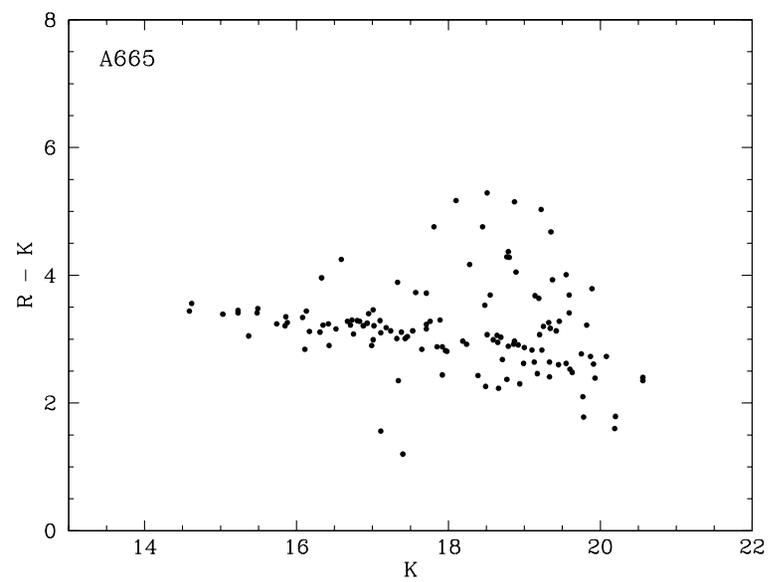
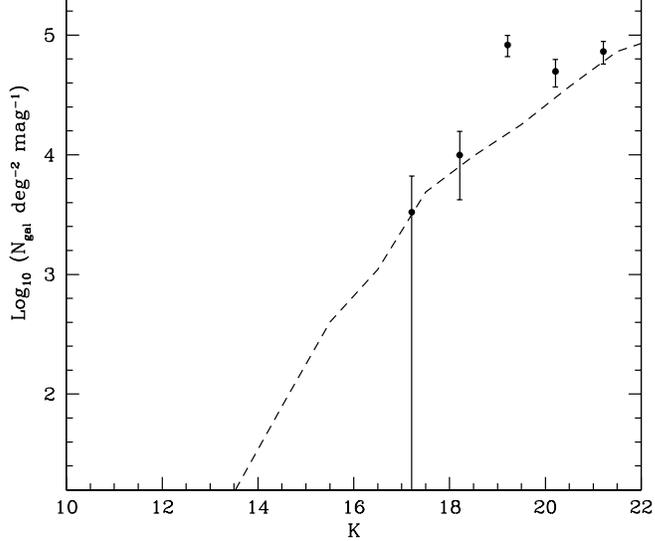
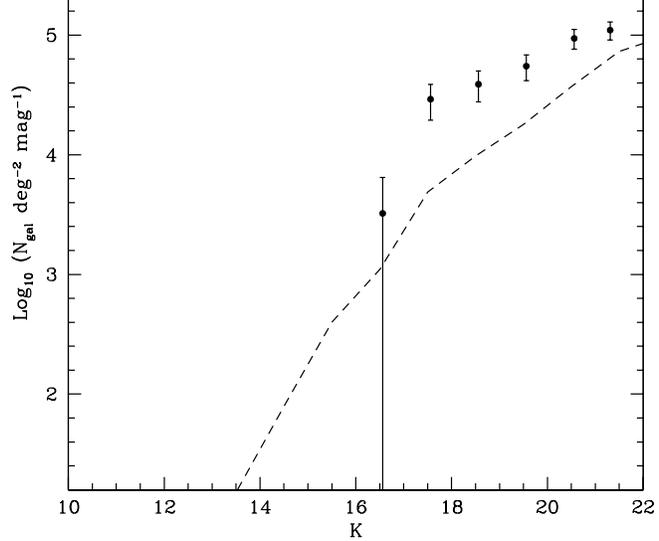
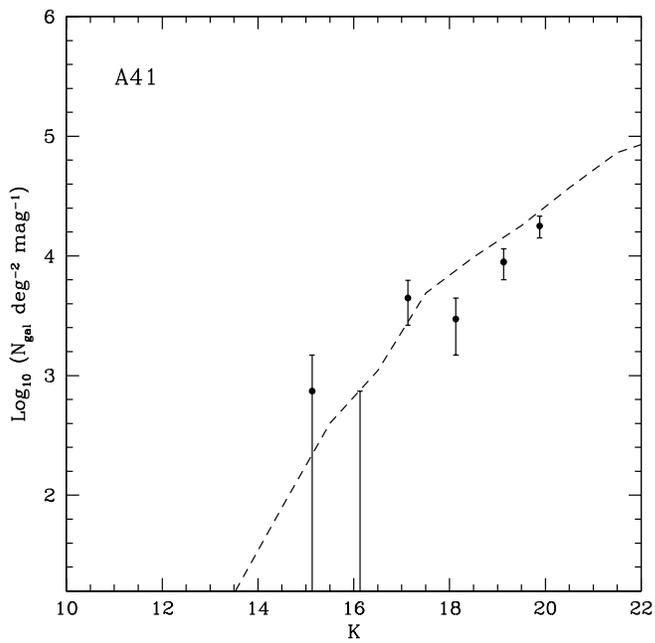
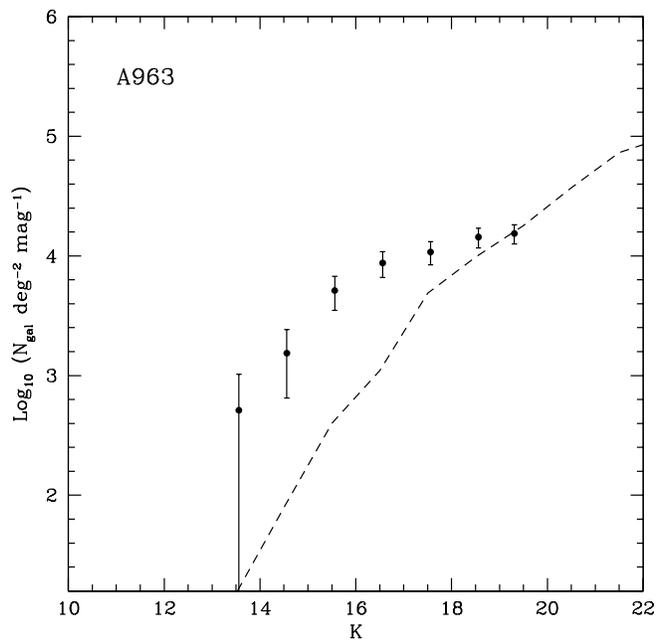
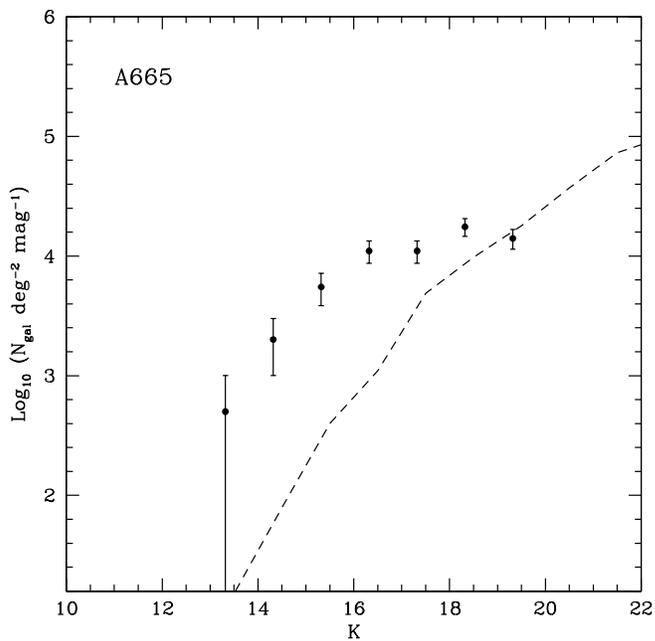
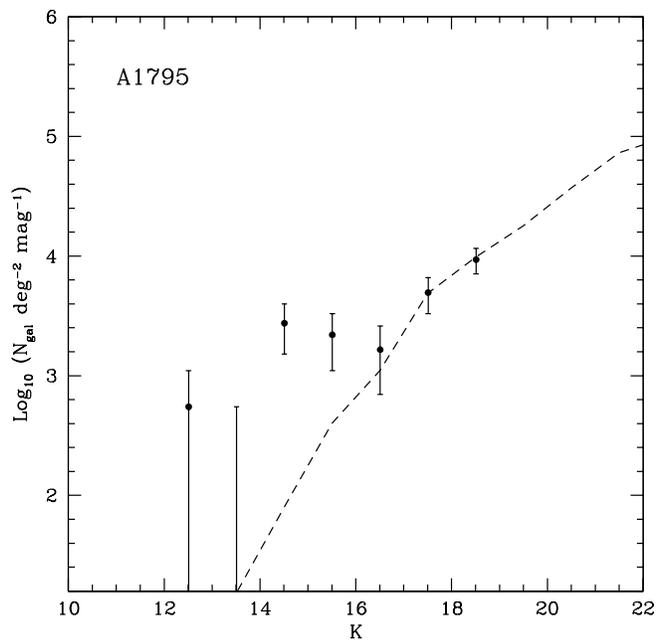
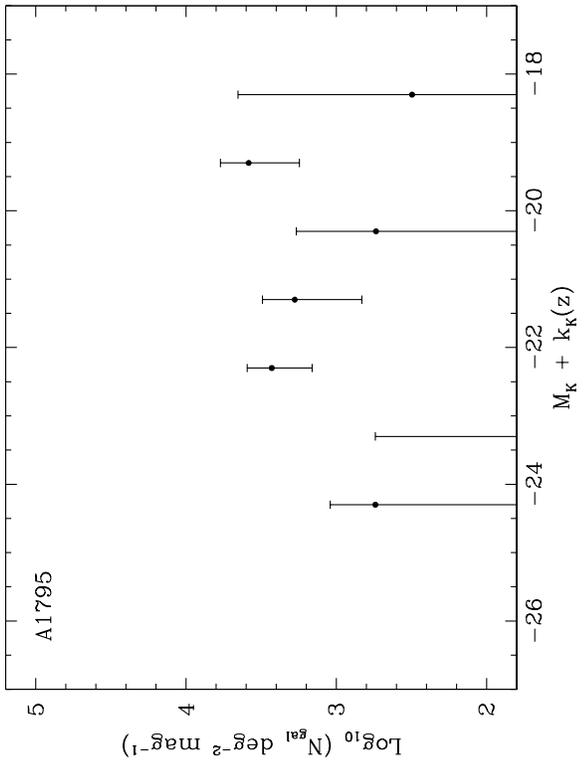
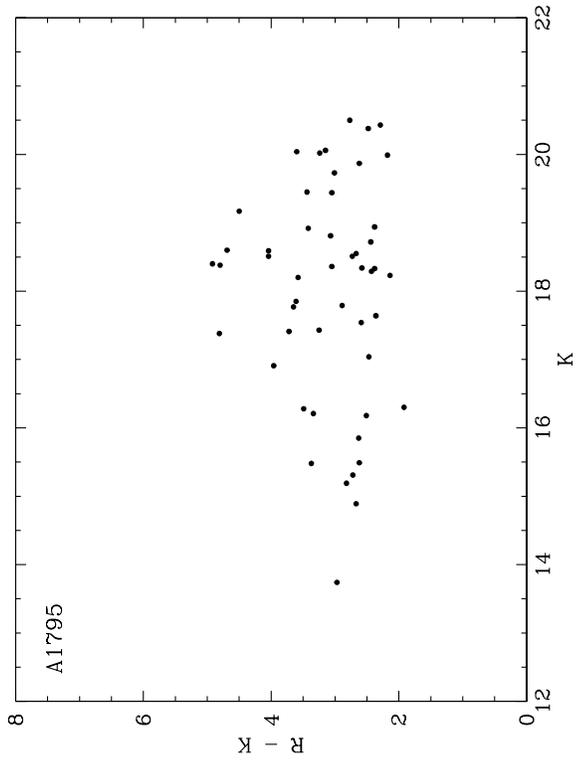
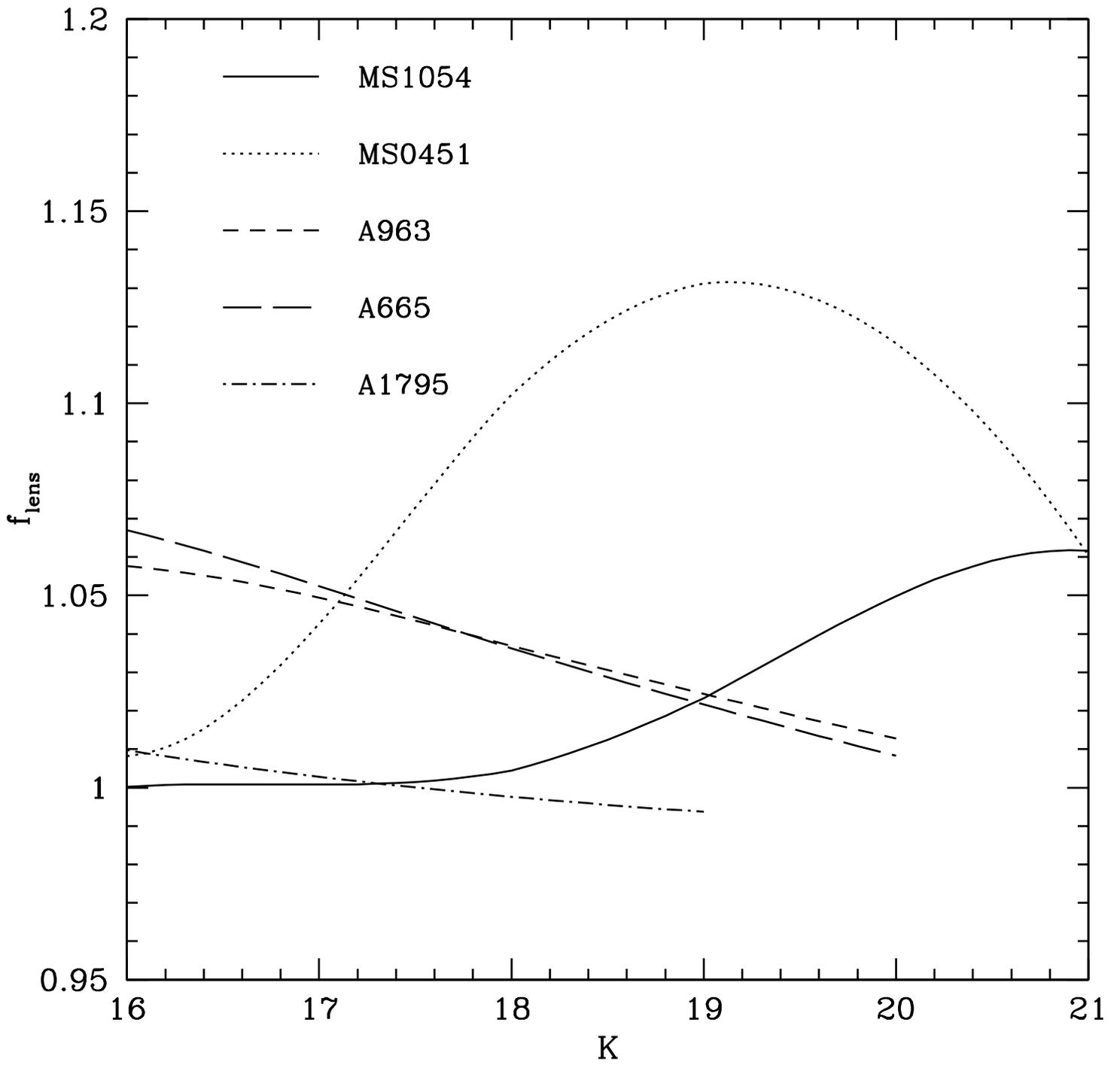
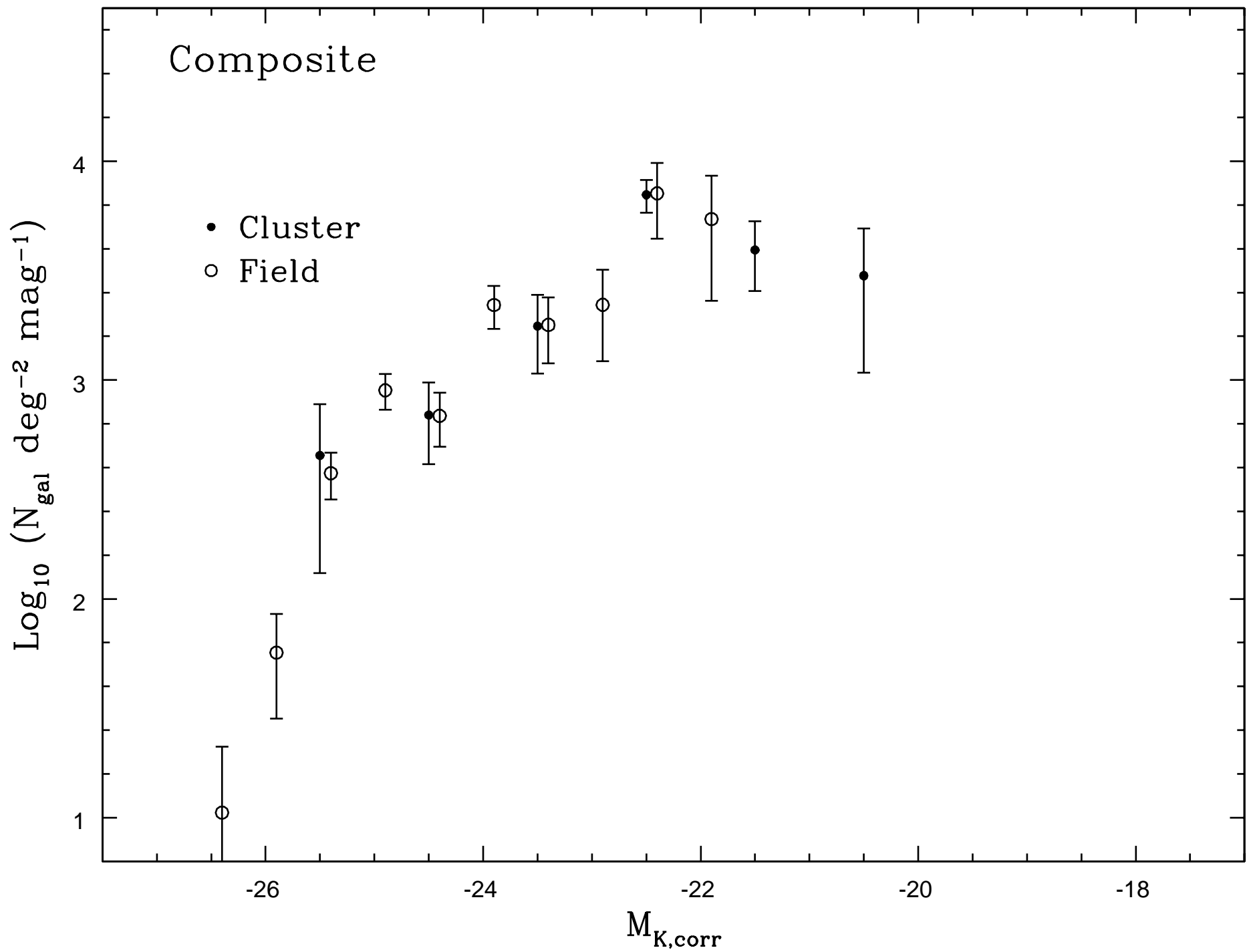
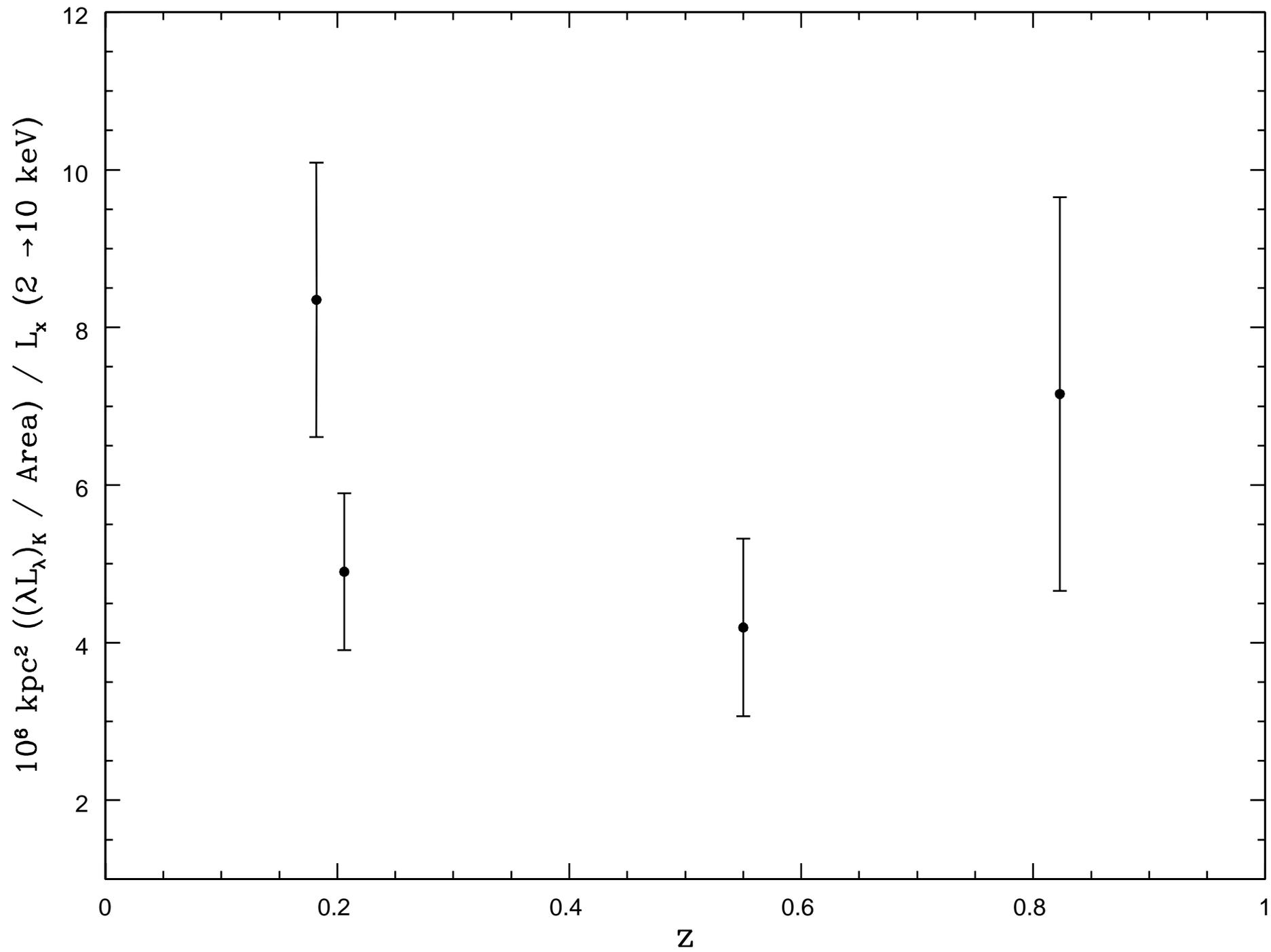
\bye